\documentclass[a4paper,10pt]{article}
\usepackage{graphicx}
\usepackage{epstopdf}
\usepackage{amsfonts,amsmath,amssymb}
\usepackage{hyperref}

\usepackage{epsfig,multicol}

\newcommand\fverb{\setbox\pippobox=\hbox\bgroup\verb}
\newcommand\fverbit{\egroup\item[\fbox{\unhbox\pippobox}]}

\newbox\pippobox

\begin{document}
\title{\bf Chameleon Gravity as an Alternative to Dark Matter }
\author{Raziyeh Zaregonbadi$^{1}$\thanks{email: r.zaregonbadi@malayeru.ac.ir}, Nasim Saba$^{2}$\thanks{email:
 n\_saba@sbu.ac.ir}\, and Mehrdad Farhoudi$^{2}$\thanks{email:
 m-farhoudi@sbu.ac.ir} \\
 {\small $^{1}$Department of Physics, Faculty of Science, Malayer University, Malayer,
 Iran}\\
 {\small $^{2}$Department of Physics, Shahid Beheshti University, 1983969411, Tehran,
 Iran}}
\date{July 18, 2025}
\maketitle
\begin{abstract}
\noindent
 By studying the chameleon gravity on galaxy scales, we
investigate the effects of the chameleon dark matter. To perform
this task, we consider the dynamics of the chameleon scalar field
in the region of galactic halos in static spherically symmetric
spacetimes that differ only slightly from classical general
relativity, but have similar symmetry behind this region. Hence,
we demonstrate its ability to act as dark matter. In fact, by
obtaining the expression for the tangential speed in this region,
we apply this approach to explain the issue of flat galactic
rotation curves. We obtain that the mass associated with the
chameleon scalar field varies linearly with the radius of galaxy,
and hence, the tangential speed in that region is constant, which
is consistent with observational data without any need to
introduce a mysterious dark matter. Accordingly, we show that the
presented chameleon gravity has a well-defined Newtonian limit and
can describe the geometry of spacetime in the region of flat
galactic rotation curves, and is consistent with the corresponding
results of the $ \Lambda CDM $ model with the NFW profile. We also
consider a test particle moving in a timelike geodesic and obtain
the fifth-force, which varies proportionally to the inverse of the
radius of galaxy. Moreover, we obtain the effects on the angle
representing the deflection of light and on the time representing
the radar echo delay of photons passing through the region of
galactic halos. We show that as the radius of galaxy increases,
the effect on the angle of light passing through the region around
galactic halos decreases. Also, the obtained result indicates that
the relevant time delay decreases with increasing radius towards
the end of the galactic halo.
\end{abstract}
\medskip
{\small \noindent
 PACS numbers:  04.50.Kd; 95.35.+d; 98.80.-k; 98.80.Jk; 04.90.+e }\newline
{\small Keywords: Chameleon Scalar Field; Dark Matter; Fifth
                  Force; Light-Deflection Angle; Radar Echo Delay }

\section{Introduction}
\indent

From various cosmological observational data, it can be said that
despite the very impressive advances of recent decades, the
unknowns of modern cosmology outweigh its knowns. About $95$
percent of the universe is made up of the mysterious unknowns of
dark energy and dark matter (DM)~\cite{p152,Planck:2018}. The
impact of dark sectors on the dynamics of the universe is well
consistent with observational data~\cite{Ishak,Ferreira}. Roughly
25$\%$ of the energy density of the universe is attributed to an
exotic matter called DM while dark energy makes up about 68$\%$.
The remaining contribution to the universe energy density belongs
to all the different materials that have been observed ever with
all instruments. Dark energy causes the accelerated expansion of
the universe, which has been supported by various observed
cosmological data~\cite{p152,Planck:2018,1}--\cite{p13}. The
effect of DM has been confirmed through the study of the rotation
curves of galaxies and the gravitational lensing, see, e.g.,
Refs.~\cite{Rubin,Clowe}. These evidences indicate the presence of
DM which does~not have the electromagnetic interaction but the
gravitational one. It has been found that the stellar
rotational/tangential speed remains constant with increasing
distance away from the galactic center. Such observations
contradict the expectation of the Newton law of gravity. It has
also been shown that galaxies have more mass than what is seen,
which is dubbed as DM. Moreover, the study of gas within
elliptical galaxies provides further evidence for DM. If the only
visible mass were the mass of galactic clusters, then they would
fall apart~\cite{binny}--\cite{Hoekstra}.

Finding a theoretical explanation for the origin of DM and dark
energy has become one of the most exciting topics during recent
decades. In this regard, many ideas have been proposed for the
form of DM, such as WIMPs (weakly interacting massive
particles)~\cite{wimp1}--\cite{wimp4}, neutralinos, sterile
neutrinos, relic neutrinos, primordial black
holes~\cite{pbh}--\cite{Fakhry2022}, axions, uncharged photinos,
and MACHOs (massive compact halo objects)~\cite{macho1,macho2}. DM
is classified into three categories: cold, warm, and hot,
depending on its speed. Cold dark matter (CDM) is heavy and slow.
It moves slowly and interacts weakly with electromagnetic
radiation and other matter. Hot DM is light and fast. Warm dark
matter (WDM) falls in between hot and cold. WIMPs, primordial
black holes and axions are some candidates for CDM, sterile
neutrinos for WDM, and relic neutrinos from the early universe for
hot DM. Among these, axions are a special case, they are very
light and extremely cold. Although CDM explains the observed
structure above $1$ Mpc well, it has problems in explaining
small-scale observations~\cite{bull} such as missing
satellites~\cite{kly,moor}, core-cusp~\cite{flor}--\cite{veg}, and
too-big-to-fail~\cite{boy,par}. To solve small-scale problems of
CDM, other candidates have been proposed, such as
WDM~\cite{blum,bod}, fuzzy DM~\cite{hu,marsh} consisting of
ultra-light bosons or scalar particles, interacting DM~\cite{spe},
and decaying DM~\cite{wang}. Other scenarios have also been
proposed that show DM comes from the model~\cite{ban}--\cite{tit}.
Nevertheless, many alternative theories have been considered that
indicate that DM effects actually arise from the theories
themselves without any need for a substance under the name
DM~\cite{Milgrom}--\cite{ZareHonar}.

In the case of dark energy, the simplest attempt is the
$\Lambda$CDM model~\cite{Deruelle}, in which the energy density is
an extraordinary small constant value. Although, this model is in
good agreement with observations, it faces problems called the
cosmological constant problem and the cosmic coincidence
problem~\cite{weincc}--\cite{Bull2016}. Among the numerous
attempts that have been made to provide a solution to such
problems, the scalar-tensor theories with a screening mechanism
are in good agreement with observations (see
Refs.~\cite{screen1}--\cite{screen2} for a review on this issue).
In the screening theories, such as
chameleon~\cite{Khoury2013,weltman1}--\cite{sabazare} and
symmetron~\cite{symetron,symetron1}, that rely on the ambient
matter density, the scalar field can act as dark energy on a
cosmic scale while remaining unseen by the local tests of gravity.

In the present work, we intend to investigate the ability of the
chameleon gravity to describe the effects of DM. For this purpose,
we consider the dynamics of the chameleon scalar field in the
region of static spherically symmetric spacetimes, and then we
attempt to show that the mass of the chameleon scalar field, in
the region of galactic halos, can play the role of DM. We have
shown in Refs.~\cite{saba,saba1} that a scalar field alone can act
as an inflaton and a chameleon and hence can cause the
acceleration of the universe at very early and late times. Now, in
the present work, we intend to show that such a scalar field can
also act as a DM. Of course, since DM itself is very difficult to
directly observe, its existence is usually detected through its
effects. One of its famous effects is the observation of flat
rotation curves of galaxies. Two other examples of these effects
are calculating the light-deflection angle~\cite{Weinberg} and the
radar echo delay~\cite{Shapiro}, all three of which we investigate
in this work. In fact, the deviation in geometry caused by DM can
affect the behavior of geodesics and it is also interesting to
study its effects.

The work is organized as follows. In  Sec.~$2$, we introduce the
desired chameleon gravity in a static spherically symmetric
spacetime and obtain the field equations of motion by taking the
variation of its corresponding action. Then, in Sec.~$3$, we
consider the relevant field equations and solutions for the
presented chameleon gravity. Thereafter, in Sec.~$4$, we first
obtain the gravitational field equations in the region of galactic
halos for solutions that differ only slightly from classical
general relativity (GR), but have similar symmetry behind this
region. Then by fixing the metric components in such a region, we
explain the resulting fifth-force by investigating the galactic
rotation curves and show that the presented chameleon gravity can
explain the observation of flat galactic rotation curves without
DM. In sections~$5$ and $6$, we investigate the propagation of
light in this presented chameleon gravity to describe the effects
of the chameleon scalar field for the light-deflection angle and
the radar echo delay. At last, we conclude the work in Sec.~$7$
with a summary of the obtained results.

\section{Chameleon Gravity in Static Spherically Symmetric System }\label{Chameleon}
\indent

The action of the chameleon gravity in four dimensions, with a
minimal coupling between the Einstein gravity and a scalar field,
say $\phi$, and a non-minimal coupling between matter fields and
this scalar field, can be given as
\begin{equation}\label{action}
S=\int {\rm d}^{4}x\left\{\sqrt{-g}\left[\frac{M^{2}_{\rm
Pl}R}{2}-\frac{1}{2}\partial_{\mu}\phi\,
\partial^{\mu}\phi-V(\phi)\right]+\sum_{i}\sqrt{-{\tilde g}^{(i)}}L^{(m)(i)}
\left({\psi}^{(m)(i)},{\tilde g}^{(i)}_{\mu\nu}\right)\right\},
\end{equation}
where $g$ is the determinant of the metric, $R$ is the Ricci
scalar, $M_{\rm Pl}=(8\pi G)^{-1/2}$ is the reduced Planck mass in
the natural units, $\hbar=1=c$, where $G$ is the gravitational
constant, and the lowercase Greek indices run from zero to three.
Also, $V(\phi)$ is a self-interacting potential, $\psi^{(m)(i)}$'s
are various matter fields, $L^{(m)(i)}$'s are the Lagrangian of
matter fields, ${\tilde g}_{\mu\nu}^{(i)}$'s are the matter field
metrics in the Jordan frame that are conformally related to the
Einstein frame metric as
\begin{equation}\label{2}
{\tilde g}^{(i)}_{\mu \nu}={e}^{ {2\frac
{{\beta}^{(i)}\phi}{M_{\rm Pl}}}}{g}_{\mu \nu},
\end{equation}
where the $\beta^{(i)}$'s are dimensionless constants representing
various non-minimal coupling between the chameleon scalar field
and each matter field. However, in this work we consider only a
single matter component and hence we omit the index $i$. In
addition, we use the conventional scalar potential for the
chameleon gravity, which must be monotonically decreasing with the
chameleon scalar field~\cite{Khoury2013}, as
\begin{equation}\label{Potential}
V\left( \phi  \right) = \frac{{{M^{4 + n}}}}{{{\phi ^n}}},
\end{equation}
where $M$ is a constant as some mass scale and $n$ is a positive
or negative integer constant.

The variation of action (\ref{action}) with respect to the metric
tensor $g_{\mu \nu}$ gives the field equations
\begin{equation}\label{G}
{G_{\mu \nu }} = \frac{1}{M_{\rm Pl}^2}\left[T_{\mu \nu
}^{(\phi)}+T_{\mu \nu }^{(m)}\right]= \frac{1}{M_{\rm
Pl}^2}\left[T_{\mu \nu }^{(\phi)}+ {e}^{ {2\frac
{{\beta}\phi}{M_{\rm Pl}}}}\widetilde {T}_{\mu\nu}^{(m)}\right],
\end{equation}
where $ T_{\mu \nu }^{(\phi)} $ is the energy-momentum tensor of
the chameleon scalar field, which is obtained to be
\begin{equation}\label{phiii}
T_{\mu \nu }^{(\phi)}  =  - \frac{1}{2}{g_{\mu \nu
}}{\partial^\alpha }\phi\, {\partial _\alpha }\phi
  - {g_{\mu \nu }}V\left( \phi  \right) + {\partial_\mu }\phi\, {\partial _\nu }\phi,
\end{equation}
and $T_{\mu\nu}^{(m)}$ and $\widetilde {T}_{\mu\nu}^{(m)}$ are the
energy-momentum tensors of matter, defined, respectively, as
\begin{equation}\label{tj}
T_{\mu\nu}^{(m)}=-\frac{2}{\sqrt{-g}}\frac{\delta\left[{\sqrt{-\tilde
{g}}}{L^{(m)}}\right]}{\delta g^{\mu\nu}}\qquad\quad {\rm and}
\qquad\quad \widetilde {T}_{\mu\nu}^{(m)}=-\frac{2}{\sqrt{-\tilde
{g}}}\frac{\delta\left[{\sqrt{-\tilde
{g}}}{L^{(m)}}\right]}{\delta {\tilde g}^{\mu\nu}},
\end{equation}
which the latter one is conserved in the Jordan frame, i.e.
${{\widetilde {\nabla}}_\mu }\widetilde {T}^{\left( m
\right)}{}_\nu^\mu  = 0$. Furthermore, the variation with respect
to the chameleon scalar field gives
\begin{equation}\label{bax}
\Box\phi=\frac{{dV\left( \phi  \right)}}{{d\phi
}}-\frac{{\beta}}{M_{\rm Pl}}e^{4\frac{{\beta}\phi}{{M}_{\rm
Pl}}}\,{\widetilde {T}}^{(m)},
\end{equation}
where $\Box\equiv\nabla^{\alpha}\nabla_{\alpha}$ corresponds to
the Einstein frame metric $g_{\mu\nu}$ and ${\widetilde
{T}}^{(m)}={\tilde g}^{\mu\nu}{\widetilde {T}}_{\mu\nu}^{(m)}$.

In this work, without loss of generality, we assume that $T^{
\left( m \right)}{}_\nu^\mu\equiv \mbox{diag}\left( { - {\rho
^{\left( m \right)}},{p^{\left( m \right)}},{p^{\left( m
\right)}},{p^{\left( m \right)}}} \right)$ and $\widetilde {T}^{
\left( m \right)}{}_\nu^\mu\equiv \mbox{diag}\left( { - {{\tilde
\rho }^{\left( m \right)}},{{\tilde p}^{\left( m
\right)}},{{\tilde p}^{\left( m \right)}},{{\tilde p}^{\left( m
\right)}}} \right)$, and also assume the matter field to be a
perfect fluid with the equation of state $p^{(m)}=w \rho^{(m)}$
and $\tilde p^{(m)}=w \tilde \rho^{(m)}$. Hence, the relation of
the matter density in the Einstein frame with the corresponding
one in the Jordan frame is
\begin{equation}\label{rhoo}
\rho^{(m)}= e^{4\frac{{\beta}\phi}{M_{\rm Pl}}}\tilde \rho^{(m)}.
\end{equation}
Moreover, we consider an isolated system described by a static and
spherically symmetric metric given by~\cite{Jackson}
\begin{equation}\label{metric}
d{s^2} =  - {e^{a\left( r \right)}}d{t^2} + {e^{b\left( r
\right)}}d{r^2} + {r^2}d{\theta ^2} + {r^2}{\sin ^2}\theta
d{\varphi ^2},
\end{equation}
where $ a(r) $ and $ b(r) $ must be found in such a way that they
satisfy the field equations. In this vein, we also assume that the
chameleon scalar field inherits the spacetime symmetry and depends
only on the radial coordinate, i.e. $\phi=\phi
(r)$\rlap.\footnote{The chameleon scalar field can depend on the
time as well, which we will consider such a case in a separate
work.}\
 Accordingly, from the conservation of ${{\widetilde {\nabla}}_\mu
}\widetilde {T}^{\left( m \right)}{}_0^\mu  = 0$, we obtain $\dot
{\tilde {\rho}}^{\left( m \right)}=0$. Also, in this case, we
have\footnote{Note that, in general, we have $0={{\widetilde
{\nabla}}_\mu }\widetilde {T}^{\left( m \right)}{}_\nu^\mu = e^{ -
4\frac{\beta \phi }{M_{\rm Pl}}}\left[ {\nabla _\mu }T^{\left( m
\right)}{}_\nu^\mu + \frac{\beta \phi' }{M_{\rm Pl}}\left( {1 -
3w} \right)\rho^{\left( m \right)}\delta^r_\nu \right]$, which
gives at least ${\nabla _\mu }T^{\left( m \right)}{}_r^\mu \neq 0$
for a general $w$ as expected.}\
\begin{equation}
0={{\widetilde {\nabla}}_\mu }\widetilde {T}^{\left( m
\right)}{}_0^\mu = e^{ - 4\frac{\beta \phi }{M_{\rm
Pl}}}\nabla_\mu T^{\left( m \right)}{}_0^\mu = -{e^{ -
4\frac{{\beta \phi }}{{{M_{\rm Pl}}}}}}{\dot \rho }^{\left( m
\right)},
\end{equation}
hence ${{\dot \rho }^{\left( m \right)}}=0$. Therefore, both the
matter densities and their pressures cannot depend on the time.
However, due to the symmetry of spacetime, we assume that those
can only depend on the radial coordinate.

We also assume that the matter in the presented chameleon gravity
is only non-relativistic baryonic matter with $ w=0 $ and that in
this study
\begin{equation}\label{PhiLess}
 \mid\beta\phi\mid \ll {M_{\rm Pl}}
\end{equation}
over the field range\rlap,\footnote{However, in the standard
chameleon gravity on the subject of dark energy, the coupling
constant parameter $\beta $ is usually considered to be of order
${\cal O}(1)$.}\
  which is imposed from cosmological and local gravity
experiments, see, e.g., Ref.~\cite{Brax2010}. The latter
assumption gives\footnote{For the Einstein frame and the Jordan
frame to be consistent with each other in the Newtonian limit, we
have to assume that the conformal factor in relation (\ref{2}) is
close to unity.}\
 $ \exp \left( {\beta \phi M_{\rm Pl}^{
- 1}} \right) \approx 1 + \beta \phi M_{\rm Pl}^{ - 1} $, which
makes contact with Brans-Dicke theories. By the way, Eq.~(\ref{G})
gives the Ricci tensor as
\begin{equation}\label{RicciT}
{R_{\mu \nu }} = \frac{1}{{M_{\rm Pl}^2}}\left[ {T_{\mu \nu
}^{\left( \phi  \right)} + T_{\mu \nu }^{\left( m \right)}}
\right]
 + \frac{1}{2}R{g_{\mu \nu }},
\end{equation}
and hence, the Ricci scalar as
\begin{equation}\label{RicciS}
R =  - \frac{1}{{M_{\rm Pl}^2}}\left[ {{T^{\left( \phi  \right)}}
+ {T^{\left( m \right)}}} \right].
\end{equation}

In addition, in the standard chameleon gravity, it is common to
define a non-physical matter density, $\rho $, as a convenient
mathematical quantity that is independent of the chameleon scalar
field. To address the subject of chameleon gravity almost
completely in the usual way, for the presented chameleon gravity,
such a quantity can also be defined as
\begin{equation}\label{physicalRho}
\rho\equiv e^{-\frac{\beta\phi}{M_{\rm Pl}}}\rho^{(m)}.
\end{equation}
Then with $ w=0 $, substituting relations~(\ref{rhoo})
and~(\ref{physicalRho}) into Eq.~(\ref{bax}) shows that the
dynamics of the chameleon scalar field is actually governed by an
effective potential, i.e.
\begin{equation}\label{field}
\Box\phi=\frac{{d{V_{\rm eff}}\left(\phi\right)}}{{d\phi }},
\end{equation}
where
\begin{equation}\label{eff}
{V}_{\rm eff}(\phi)\equiv V(\phi)+\rho\,
e^{\frac{\beta\phi}{M_{\rm Pl}}}.
\end{equation}
However, in the presented chameleon gravity, such the non-physical
matter density is~not related to a conserved energy-momentum
tensor in the Einstein frame (see Eq.~(\ref{rhoEquation}) in the
following), while it is related in the standard chameleon gravity.

\section{Relevant Field Equations and Solutions for the Model}
\indent

In this section, we consider the field equations for metric
(\ref{metric}) to obtain the metric components for the presented
chameleon gravity under consideration. In this regard, inserting
metric (\ref{metric}) into the field equations~(\ref{G}) gives
\begin{equation}\label{field eq1}
G_t^t =  - \frac{1}{{{r^2}}} + \frac{1}{{{r^2}{e^b}}} -
\frac{{b'}}{{r{e^b}}} =-\frac{1}{{M_{\rm Pl}^2}}\left[ {\rho
^{\left( \phi  \right)}}+
 \rho^{\left( m \right)}\right],
\end{equation}
\begin{equation}\label{field eq2}
G_r^r =  - \frac{1}{{{r^2}}} + \frac{1}{{{r^2}{e^b}}} +
\frac{{a'}}{{r{e^b}}} =\frac{p_r^{\left( \phi \right)}}{{M_{\rm
Pl}^2}}
\end{equation}
and
\begin{equation}\label{field eq3}
G_\theta ^\theta  = G_\varphi ^\varphi = \frac{1}{{4{e^b}}}\left(
{2a'' + {{a'}^2} - a'b'} \right) + \frac{{a' - b'}}{{2r{e^b}}}
=\frac{p_ \bot ^{\left( \phi \right)} }{{M_{\rm Pl}^2}},
\end{equation}
where the prime represents derivative with respect to the radial
coordinate. Also, we have used $ T^{\left(\phi \right)}{}_\nu^\mu
\equiv \mbox{diag}\left(- {\rho^{\left( \phi \right)}},p_r^{\left(
\phi \right)},p_ \bot ^{\left(\phi \right)},p_ \bot ^{\left( \phi
\right)} \right) $, where $ {p_r^{\left( \phi \right)}} $ and $
{p_ \bot ^{\left( \phi \right)} } $  correspond to the radial and
tangential pressure components of the energy-momentum tensor of
the chameleon scalar field, respectively. However, from relation
(\ref{phiii}) while using metric (\ref{metric}), we obtain
\begin{equation}\label{density}
{\rho ^{\left( \phi \right)}} =  - T^{\left( \phi \right)}{}_t^t =
\frac{{{{\phi '}^2}}}{{2{e^{b\left( r \right)}}}} + V\left( \phi
\right),
\end{equation}
\begin{equation}\label{press}
p_r^{( \phi)} = T^{\left( \phi \right)}{}_r^r= \frac{{{{\phi
'}^2}}}{{2{e^{b\left( r \right)}}}} - V\left( \phi  \right),
\end{equation}
\begin{equation}\label{pressteta}
p_ \bot ^{\left( \phi  \right)} = T^{\left( \phi
\right)}{}_\theta^\theta = T^{ \left( \phi
\right)}{}_\varphi^\varphi=-\frac{{{{\phi '}^2}}}{{2{e^{b\left( r
\right)}}}} - V\left( \phi \right).
\end{equation}
These relations indicate that, in accordance with the symmetry of
spacetime, these components also ultimately depend on the radial
coordinate. Moreover, from relation (\ref{field eq1}), we have
\begin{equation}
\int {\left( {\frac{1}{{{e^b}}} - \frac{{b'r}}{{{e^b}}}} \right)}
dr = \int {\left[ {1 - \frac{{{r^2}}}{{M_{\rm Pl}^2}} \left(
{{\rho ^{\left( \phi  \right)}} + \rho^{\left( m \right)} }
\right)} \right]} dr,
\end{equation}
that gives
\begin{equation}
\frac{r}{{{e^b}}} = r - \int \frac{{{r^2}}}{{M_{\rm Pl}^2}}\left(
{{\rho ^{\left( \phi  \right)}}
 + \rho^{\left( m \right)} } \right)dr.
\end{equation}
Hence, we get
\begin{eqnarray}\label{e^-b}
{e^{ - b\left( r \right)}} &=& 1 - \frac{1}{{4\pi\, r
M_{{\rm{Pl}}}^2}}\int {4\pi \left( {{\rho ^{\left( \phi \right)}}
 + \rho^{\left( m \right)} } \right)} {{r}^2}dr\cr
 & =& 1 - \frac{1}{{4\pi\, r M_{{\rm{Pl}}}^2}}\left( {{M^{\left( \phi  \right)}} + {M^{\left( m \right)}}}
\right)=1-\frac{2G M^{\rm \left(tot \right)}(r)}{r}
\end{eqnarray}
that is reminiscent of the Newtonian limit of a point mass
situated at the origin. Here, $M^{\left( \phi \right)}$ and
$M^{\left( m \right)}$ are the mass associated with the chameleon
scalar field and the baryonic mass, respectively, and we have
considered the total mass of the object in question (for example a
typical galaxy/galactic cluster) to be $ {M^{\rm \left(tot
\right)}} \equiv {{M^{\left( \phi \right)}} + {M^{\left( m
\right)}}} $.

In the following, we derive a useful relation from metric
(\ref{metric}) as
\begin{equation}\label{R,a,b}
\frac{{{R_{tt}}}}{{{e^{a\left( r \right)}}}} +
\frac{{{R_{rr}}}}{{{e^{b\left( r \right)}}}} = \frac{{a' +
b'}}{{r{e^{b\left( r \right)}}}}
\end{equation}
to obtain the components of the metric in the presented chameleon
gravity. In this regard, from relations (\ref{RicciT}) and
(\ref{RicciS}), we have
\begin{equation}
{R_{tt}} = \frac{{{e^{a\left( r \right)}}}}{{M_{\rm Pl}^2}}
 \left[ {{\rho ^{\left( \phi  \right)}}+\rho^{\left( m \right)}
 }\right]
 - \frac{{{e^{a\left( r \right)}}}}{2}R
\end{equation}
and
\begin{equation}
{R_{rr}} = \frac{{{e^{b\left( r \right)}}}}{{M_{\rm
Pl}^2}}p_r^{\left( \phi  \right)} + \frac{{{e^{b\left( r
\right)}}}}{2}R.
\end{equation}
Hence, we obtain
\begin{equation}\label{R,R}
\frac{{{R_{tt}}}}{{{e^{a\left( r \right)}}}} +
\frac{{{R_{rr}}}}{{{e^{b\left( r \right)}}}} = \frac{1}{{M_{\rm
Pl}^2}}\left[ {
 {\rho^{\left( m \right)}
 +{\rho ^{\left( \phi  \right)}} }+ p_r^{\left( \phi  \right)}}
 \right].
\end{equation}

\section{Geometry Within Galactic Halo}
\indent

Observational data show that as one moves away from the core of a
typical galaxy/galactic cluster up to several luminous radii, the
tangential speed increases linearly within the bulge, approaching
a constant value around $ 200-500\ km/s $~\cite{binny,Persic}. In
the region of galactic halos, where the baryonic matter density $
\rho^{\left( m \right)} $ is very low, we can plausibly assume $
\rho^{\left( m \right)} \ll {\rho ^{\left( \phi \right)}} $ and
hence, neglect the contribution of the baryonic matter compared to
the mass associated with the chameleon scalar
field\rlap.\footnote{This assumption prevents the direct effect of
the chameleon coupling parameter $\beta$ on the chameleon scalar
field resulting from the field equations~(\ref{G}), however see
the continuation.}\
 Then, by inserting relations~(\ref{density})
and (\ref{press}) into relation~(\ref{R,R}), and using relation
(\ref{R,a,b}), we obtain
\begin{equation}\label{a,b}
\frac{{a' + b'}}{r} \approx  \frac{{{\phi '}^2}}{{M_{\rm Pl}^2}}.
\end{equation}

It is known that the gravitational field outside the halo is
represented by the exterior Schwarzschild metric and inside the
halo by $ {e^{a\left( r \right)}} $ and $ {e^{b\left( r \right)}}
$ in metric~(\ref{metric}), which can be determined from
observations of the rotation curves. For the presented chameleon
gravity, we intend to achieve solutions in the region of galactic
halos that differ only slightly from classical GR, but have its
symmetry behind this region, namely the symmetry $e^{a\left( r
\right)}=e^{-b\left(r \right)}$ with relation
(\ref{e^-b})\rlap.\footnote{Behind the region of galactic halos,
as is obvious from relation (\ref{e^-b}), this solution is
asymptotically flat.}\
 For this purpose, we first assume
\begin{equation}\label{FirstAssume}
{e^{a\left( r \right)}}{e^{b\left(r \right)}} = F\left( r \right)
,
\end{equation}
and to remain in the vicinity of GR, such a function $ F(r) $
needs to differ slightly from $ 1 $. Then, to achieve such a goal,
we consider
\begin{equation}\label{Fofr}
F\left( r \right) = {\left( {\frac{r}{s}} \right)^\alpha },
\end{equation}
where the parameter $ s $ is a length scale of the system, which
we consider as the visible radius of the galaxy (the radius of
baryonic matter) beyond the edge of the stellar
disk\rlap.\footnote{The visible radius is the radius at which the
light (from stars or emitting gas) of a galaxy is detectable. In
contrast, the edge of the stellar disk is the radius at which the
surface brightness of the starlight drops sharply or the
exponential structure of the stellar disk is truncated. That is,
the edge of the stellar disk is usually smaller and only
corresponds to starlight, while the visible radius can also
include gaslight.}\
 As an example, we have plotted these radii for the NGC~$5533$
galaxy with data from the SPARC database in Figure~$(1)$.
\begin{figure}[h]
\begin{center}
\includegraphics[scale=0.35]{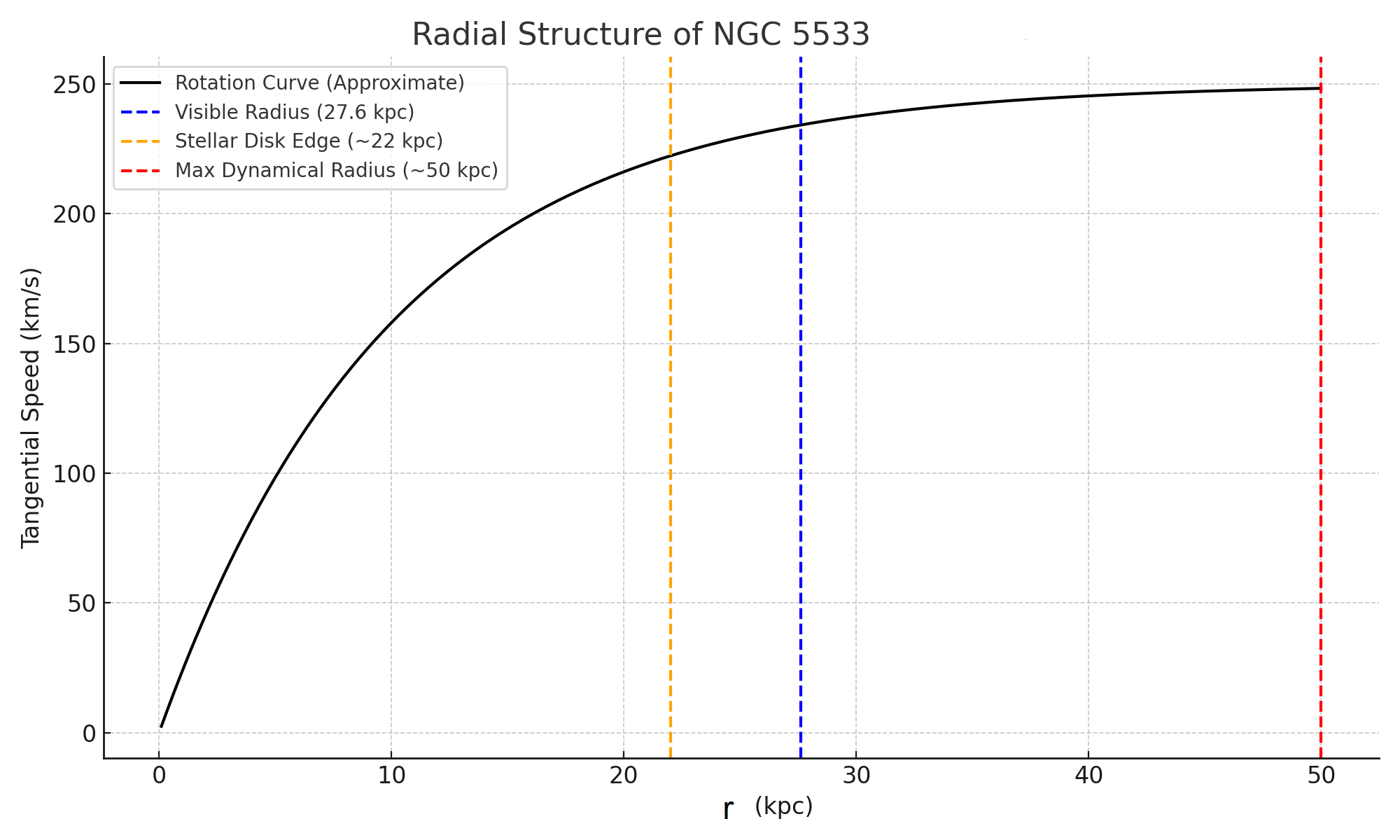}
\caption{The figure illustrates the visible radius and the stellar
disk edge for the NGC~$5533$ galaxy with data from the SPARC
database.}
\end{center}
\end{figure}
Also, the parameter $ \alpha $ is a positive dimensionless
parameter that must be $\alpha\ll 1$, but not~zero. In this
regard, by alternatively writing $F(r)$ in relation~(\ref{Fofr})
as $F\left( r \right) = \exp \left[ {\ln {{\left( {r/s}
\right)}^\alpha }} \right]$, we easily obtain
\begin{equation}
F\left( r \right) \approx 1 + \alpha \ln \left( {\frac{r}{s}}
\right),
\end{equation}
as required. Now, by taking the derivative of
relation~(\ref{FirstAssume}) while using relation~(\ref{Fofr}), we
get
\begin{equation}\label{Fofr2}
a' + b' = \frac{\alpha }{r}.
\end{equation}
Thereafter, by inserting relation (\ref{Fofr2}) into
relation~(\ref{a,b}), we obtain
\begin{equation}\label{phi prime}
\phi '  \approx  \pm \frac{{{M_{\rm Pl}}{\sqrt{\alpha}}}}{r},
\end{equation}
which gives
\begin{equation}
\phi \left( r \right)- \phi \left( {{s}} \right)
 \approx  \pm {M_{\rm Pl}}{\sqrt{\alpha}}\ln \left( {\frac{r}{s}}
 \right).
\end{equation}
If we plausibly assume that the chameleon scalar field is zero on
the boundary radius $s$ of the baryonic matter, then in the region
of galactic halos, $ \phi \left( r \right) $ will be
\begin{equation}\label{phiR}
\phi \left( r \right)\approx\pm {M_{\rm Pl}}\sqrt{\alpha}\ln
{\left( {\frac{r}{s}} \right) },
\end{equation}
which obviously yields $\phi \left( {{s}} \right)=0$ as
required\rlap.\footnote{Also, at the radius $s$, from relation
(\ref{2}), we obviously get $ \tilde{g}_{\mu\nu}|_{r=s}
=g_{\mu\nu}$.}\
 We recall that relation~(\ref{phiR}) is for $\alpha \neq 0$,
otherwise for $\alpha = 0$ behind the region of galactic halos,
$\phi'=0$ and the chameleon scalar field is constant.

Furthermore, we consider that the baryonic matter dominates in the
interior of galaxies up to $ r=s $ and, in comparison, we
plausibly assume that the non-local effects of the chameleon
scalar field can be neglected. Hence, from relation (\ref{e^-b}),
the metric component $ g_{rr} $ at the boundary of the baryonic
matter with the galactic halo is
\begin{equation}\label{ebs}
{e^{ - b\left( s \right)}} \approx 1 - \frac{{{\left. {{M^{\left(
m \right)}}} \right|}_{r = s}}}{{4\pi s\, M_{\rm Pl}^2}},
\end{equation}
where the mass of baryonic matter at the $  r=s $ boundary,
$M^{\left( m \right)}\big|_{r = s}$, has a certain constant value
equal to the observable mass in a typical galaxy/galactic cluster.
On the other hand, in the region of galactic halos, i.e., in the
range\footnote{The parameter $ r_{\rm D}^{} $ is the radius at
which the halo terminates, i.e., up to where the rotation curve is
flat. It is also considered as the radius of DM, behind which the
contribution of all masses vanish.}\
  $ s\leq r\leq r_{\rm D}^{} $, the total mass can be estimated as
the mass associated with the chameleon scalar field, hence
relation (\ref{e^-b}) reads
\begin{equation}\label{ebb}
{e^{ - b\left( r \right)}} \approx 1 - \frac{{{M^{\left( \phi
\right)}}}}{{4\pi r\, M_{\rm Pl}^2}}.
\end{equation}
Now, the continuity of the metric coefficient across the radius of
the region dominated by the baryonic matter leads to relation
(\ref{ebb}) being proportional to relation (\ref{ebs}), which
gives
\begin{equation}\label{M phi r}
{M^{\left( \phi  \right)}} \propto \left( {\frac{{{{\left.
{{M^{\left( m \right)}}} \right|}_{r = s}}}}{s}} \right)r.
\end{equation}
Therefore, in the region of galactic halos, we obtain that the
mass associated with the chameleon scalar field, $ {M^{\left( \phi
\right)}} $, varies linearly with the radius of galaxy/galactic
cluster, $ r $. In this regard, according to the Newtonian
gravity~\cite{binny}, the galaxy rotation curves show that the
speed of matter in a spiral disk rotates as a function of the
radius, i.e., $v_{\rm tg}^{} \left(r\right) = \sqrt {G{M^{\rm
\left(tot \right)}}/r}$ in the natural units, which is
experimentally constant in the halo of galaxies. Since the
baryonic matter is negligible in the region of flat galactic
rotation curves, the total mass is essentially determined by the
mass associated with the chameleon scalar field. That is, the
chameleon scalar field plays a crucial role in assessing the total
mass as a dominant factor in this region. Hence, by considering
the chameleon scalar field, in the presented chameleon gravity, we
can accurately achieve
\begin{equation}\label{speed}
v_{\rm tg}^{} \approx \sqrt {\frac{{G{M^{\left( \phi
\right)}}}}{r}}={\rm const.},
\end{equation}
where we have utilized relation (\ref{M phi r}) in the last term.
This result indicates that the tangential speed is a constant
value in the halo of galaxies. Accordingly, we conclude that the
mass associated with the chameleon scalar field can explain the
observation of flat galactic rotation curves without any need to
introduce a mysterious DM. That is, the mass associated with the
chameleon scalar field plays the role of the mass associated with
DM. Indeed, amidst the extensive efforts to search for particle DM
candidates, the presented chameleon gravity represents a different
approach. In other words, since finding the nature and properties
of DM is one of the outstanding questions in contemporary
cosmology, such a result could be interesting and demonstrate the
capability of the presented chameleon gravity. Nevertheless, to
account for the variety of behaviors of observed rotation curves
in full spectrum and empirical relations like the baryonic
Tully-Fisher relation, a possible approach in this presented
chameleon gravity could be to consider the parameter $s$ dependent
on the radius $r$ and, in turn, the continuity of the metric
coefficient across the corresponding radius, which we will explore
in future work.

Moreover, by substituting relation (\ref{M phi r}) into relation
(\ref{ebb}), it is shown that the $ {g_{rr}} $ component of the
metric is constant in the region of galactic halos. Accordingly,
inserting $b' = 0$ of this region into relation (\ref{Fofr2})
yields the $ g_{tt} $ component as
\begin{equation}\label{gtt}
da = \alpha \frac{{dr}}{r}\qquad \Rightarrow \qquad {{e^{a\left( r
\right)}}}
 = {\left( {\frac{r}{R_0}} \right)^\alpha }
\end{equation}
in the region of galactic halos with $\alpha \neq 0$, where $
{R_0} $ is a constant of integration with positive values. Also,
by substituting relation (\ref{gtt}) into relation
(\ref{FirstAssume}) while employing relation (\ref{Fofr}), we
achieve the constant value of the $ g_{rr} $ component to be
\begin{equation}\label{grr}
{e^{b}} = {\left( {\frac{{{R_0}}}{s}} \right)^\alpha },
\end{equation}
in the region of galactic halos with $\alpha \neq 0$.

Before proceeding, let us mention that if one chooses the constant
value of $R_0$ to be $R_0=s$, one will have $e^b=1$. Then, at the
radius $s$, one also gets $e^{a(s)}=1$. Therefore, at this
boundary radius, where the chameleon scalar field also does~not
exist, the metric becomes the Minkowski metric. However, we remind
that although $a(s)=0$, but $a'(s)=\alpha/s\neq 0$. On the other
hand, by inserting relation (\ref{grr}) into relation (\ref{ebb})
while using relation (\ref{M phi r}), we obtain the constant value
\begin{equation}\label{MmAt-s}
M^{\left( m \right)}|_{r = s}\propto 4\pi\, s\, \alpha M^2_{\rm
Pl}\ln(R_0/s),
\end{equation}
wherein $\alpha \neq 0$. Accordingly, this result dictates that
one cannot choose  $R_0=s$, but rather $R_0>s$ must be. In
addition, inserting relation~(\ref{MmAt-s}) into
relation~(\ref{speed}) yields
\begin{equation}\label{speed2}
v_{\rm tg}^{} \propto \sqrt {\frac{\alpha\ln(R_0/s)}{2}}
\end{equation}
in the region of galactic halos with $\alpha \neq 0$. Now,
utilizing the experimental data of each galaxy/galactic cluster
for $s$, $v_{\rm tg}^{}$ and $M^{\left( m \right)}|_{r = s}$,
through relations~(\ref{MmAt-s}) and~(\ref{speed2}), we can
determine the two constant parameters $R_0$ and $\alpha$ of the
presented chameleon gravity for that galaxy/galactic cluster. In
this case, there will be a set of different parameters
($R_0,\alpha$). Moreover, substituting
relations~(\ref{Potential}), (\ref{phi prime}), (\ref{phiR})
and~(\ref{grr}) into relation~(\ref{density}) yields the mass
density of the chameleon scalar field as the mass density of DM
for each galaxy/galactic cluster as
\begin{equation}\label{density2}
{\rho ^{\left( \phi \right)}}\approx \frac{\alpha M^2_{\rm
Pl}(s/R_0)^{\alpha} }{2 r^2}+\frac{(\pm)^{n} M^{4+n}}{M^{n}_{\rm
Pl}\,\alpha^{n/2}\left[\ln\left(\frac{r}{s}\right)\right]^{n}},
\end{equation}
where its positive values are acceptable, which we will discuss
below relation~\eqref{rhoAs-r}. Also, in the following, instead of
relation~\eqref{density2}, we will derive a more practical
relation for $\rho^{\left( \phi \right)}$ (i.e.,
relation~\eqref{new rho phi}).

Meanwhile, from the conservation of ${{\widetilde {\nabla}}_\mu
}\widetilde {T}^{\left( m \right)}{}_1^\mu  = 0$, we have
\begin{equation}\label{rhoPrime}
w\tilde{\rho}^{\left(m\right)'}+\frac{1}{2}\tilde{a}'(r)(1+w)\tilde{\rho}^{\left(
m \right)}=0,
\end{equation}
where $\tilde{a}(r)\equiv a(r) +2\beta \phi(r)/M_{\rm
Pl}$\rlap.\footnote{The corresponding equations for $\rho^{(m)}$
and the non-physical matter density $\rho $ are
\begin{equation}\label{usualRho}
w\rho^{\left(m\right)'}+\frac{1}{2}a'(r)(1+w)\rho^{\left( m
\right)}=-\frac{\beta\phi'}{M_{\rm
Pl}}(1-3w)\rho^{\left(m\right)},
\end{equation}
\begin{equation}\label{rhoEquation}
w\rho'+\frac{1}{2}a'(r)(1+w)\rho=-\frac{\beta\phi'}{M_{\rm
Pl}}(1-2w)\rho.
\end{equation}
}\
 In the case $w=0$, by substituting
relations~(\ref{phi prime}) and~(\ref{gtt}) into
relation~(\ref{rhoPrime}), we obtain
\begin{equation}\label{alphaBeta}
\beta \approx\mp\frac{\sqrt{\alpha}}{2}\qquad\qquad {\rm
and/or}\qquad\qquad \alpha \approx 4\beta^2,
\end{equation}
in the region of galactic halos with $\alpha \neq 0$. It is
important to note that for $\alpha = 0$, there is no~relation
between $\beta$ and $\alpha$. Since $0<\alpha\ll 1$, the above
result indicates that, in the region of galactic halos, the
chameleon coupling parameter is committed to be $-1\ll\beta\ll 1$
in the presented chameleon gravity as an alternative to dark
matter. The obtained upper limit is well consistent with the
expected magnitude and possible observational bounds on the
chameleon coupling parameter, see, e.g.,
Refs.~{\cite{Brax2010,Mota2011,Boriero}. Meanwhile, by inserting
relations~(\ref{phiR}) and~(\ref{alphaBeta}) into
assumption~(\ref{PhiLess}), we obtain
\begin{equation}\label{NewCondition}
 2\beta^2\ln
{\left( {\frac{r}{s}} \right) } \ll 1.
\end{equation}
Even for $r_{\rm D}^{}$ as the maximum $r$ with its value in a
typical galaxy/galactic cluster, say $r_{\rm D}^{}\approx e^{2}
s$, condition~\eqref{NewCondition} gives a value of $\beta $ much
less than $1$. Hence, at least in the region of galactic halos,
the restrictive assumption~(\ref{PhiLess}) is fulfilled.

Furthermore, using metric~(\ref{metric}) for $\Box\phi(r)$, it
gives
\begin{equation}\label{boxPhi}
\Box\phi(r)=e^{-b}\left[\phi''+\frac{1}{2}\left(a'-b'+\frac{4}{r}\right)\phi'\right].
\end{equation}
Then, by putting this relation into the field equation~(\ref{bax})
while knowing that $b'=0$ in the region of galactic halos and
utilizing relations~(\ref{Potential}), (\ref{rhoo}), (\ref{phi
prime}), (\ref{phiR}), (\ref{gtt}), (\ref{grr})
and~(\ref{alphaBeta}), we get $\rho^{(m)}$ in the region of
galactic halos as a function of $r$ as
\begin{equation}\label{rhoAs-r}
\rho^{(m)}(r)\approx -\frac{M^2_{\rm Pl}(s/R_0)^{\alpha}
(2+\alpha)}{r^2}\mp\frac{(\pm)^{n+1}\, 2nM^{4+n}}{M^{n}_{\rm
Pl}\,\alpha^{n/2+1}\left[\ln\left(\frac{r}{s}\right)\right]^{n+1}},
\end{equation}
where $\alpha \neq 0$. We recall that for $\alpha =0$ behind the
region of galactic halos, since $\phi$ is constant and
relation~(\ref{boxPhi}) vanishes, the field equation~(\ref{bax})
yields $\rho^{(m)}$ to be zero as expected. Meanwhile, for
$\rho^{(m)}$ to have a positive value, since $r>s$, relation
(\ref{rhoAs-r}) dictates that only odd values of $n$ with the
lower-sign of the second fraction can be solutions. That is, the
power $n$ of the chameleon scalar field in the corresponding
potential~(\ref{Potential}) must be odd values, and in the results
obtained, including relations~(\ref{phi prime}), (\ref{phiR})
and~(\ref{alphaBeta}), the lower-sign is the acceptable sign.
Explicitly, in the region of galactic halos, the parameter $\beta$
must have positive values, although much less than $1$. In
addition to the fact that $\rho^{(m)}$ in relation (\ref{rhoAs-r})
must be positive, $\rho ^{\left( \phi \right)}$ in
relation~(\ref{density2}) must also be positive, and the
contribution of the baryonic matter compared to the mass
associated with the chameleon scalar field must also be neglected
in the region of galactic halos. Under these considerations, by
neglecting the term $\alpha$ compared to the term $2$ in the first
term on the right-hand side of relation (\ref{rhoAs-r}) and
assuming $r=e^x\, s$, e.g. for $n=1$, after some calculations, all
the conditions when $0<x<1/2$ are fulfilled.

Alternatively, the field equation~(\ref{bax}) can be used to
obtain another suitable potential instead of the standard
chameleon potential~(\ref{Potential}). In this case, by neglecting
the expression containing $\rho^{(m)}$ and utilizing other
relevant relations, we obtain
\begin{equation}
V(r)\approx -M^2_{\rm Pl}\,\alpha
\left(1+\frac{\alpha}{2}\right)\left(\frac{s}{R_0}\right)^{\alpha}\frac{1}{2\,
r^2}.
\end{equation}
However, we do~not pursue this result in this work.

In the continuation, it is well-known that in the Newtonian limit
the $ g_{tt} $ component of the metric is given by $ {e^a} \approx
1 + 2{\Phi _{\rm N}} $, where $ {\Phi _{\rm N}} $ is the Newtonian
gravitational potential satisfying the Poisson
equation~\cite{Landau}. In the constant speed region, this
potential has a general solution $ {\Phi _{\rm N}} \approx v_{\rm
tg}^2\ln r $, where $ v_{\rm tg}^2 \approx
{\mathcal{O}}{(10^{-6})}$ is the tangential speed of a test
particle in the natural units in the region of galactic
halos~\cite{Bohmer}. In the presented chameleon gravity, by
alternatively writing the $ g_{tt} $ component in relation
(\ref{gtt}) as $e^{a\left( r \right)} = \exp \left[ \ln \left(
r/R_0\right)^\alpha \right]$, we can easily obtain
\begin{equation}
{e^{a\left( r \right)}} \approx 1 + \alpha \ln \left(
{\frac{r}{{{R_0}}}} \right)
\end{equation}
in the region of flat galactic rotation curves. This result shows
that the presented chameleon gravity has a well-defined Newtonian
limit and can describe the geometry of spacetime in this region.
In addition, as mentioned around relation (\ref{Fofr}), in order
to remain close to GR, the parameter $ \alpha $ must be much less
than $1$, hence, in comparison with the Newtonian limit (up to a
normalization factor), we plausibly assume that the value of $
\alpha $ is approximately the value of $v_{\rm tg}^2$ in the
natural units. Hence, by considering relation~(\ref{speed2}), the
parameter $R_0$ is
\begin{equation}\label{R0}
\ln(R_0/s)\propto 2\qquad \Rightarrow \qquad R_0\propto e^2\, s.
\end{equation}
In this way, we have chosen the two constant parameters
($R_0,\alpha$) to be used for all galaxies/galactic clusters, and
from now on we will use these chosen values instead. Indeed, in
this way, we have somehow circumvented the point that the
fundamental equations of the presented chameleon gravity, and
especially the chameleon coupling parameter, vary from galaxy to
galaxy\rlap.\footnote{However, for considering a field-dependent
coupling parameter in action, see, e.g.,
Refs.~\cite{Brax2010,Mota2011}.}\
 Instead, the predicted
results from the presented chameleon gravity depend on the data
for each galaxy, as can be seen in the results of the following
sections. Also, for instance, the mass of baryonic matter at the
boundary, relation~(\ref{MmAt-s}), now becomes $M^{\left( m
\right)}|_{r = s}\propto 8\pi\, s\, v_{\rm tg}^2 M^2_{\rm Pl}= s\,
v_{\rm tg}^2/G$, which recalls relation~(\ref{speed}).

To proceed further, we consider a test particle moving in a
timelike geodesic in the static and spherically symmetric system
in the Einstein frame. In this regard, the Newtonian limit of the
geodesic equation is~\cite{Burrage,Faraoni}
\begin{equation}
\frac{{{d^2}{x^\mu }}}{{d{\tau ^2}}} + \Gamma^\mu{}_{\nu \sigma }
\frac{{d{x^\nu }}}{{d\tau }}\frac{{d{x^\sigma }}}{{d\tau }} = -
\frac{v_{\rm tg}^{}} {{{2M_{\rm Pl}}}} {\nabla ^\mu }\phi,
\end{equation}
where $ \tau $ is an affine parameter along the geodesics of
$g_{\mu \nu } $ and $\Gamma^\mu{}_{\nu \sigma }$ is the
Christoffel symbol. This equation illustrates that there is a
fifth-force proportional to the gradient $ \phi $ that couples to
any massive test particle. Accordingly, the corresponding geodesic
equation for the $ r $-coordinate in the Einstein frame is
\begin{equation}\label{geodesic}
\frac{{{d^2}r}}{{d{\tau ^2}}} + \Gamma^r{}_{tt} =
 - \frac{v_{\rm tg}^{}} {{{2M_{\rm Pl}}}}{\nabla ^r}\phi.
\end{equation}
Therefore, for stable circular orbits, the Newtonian force also
contains~\cite{Burrage,Faraoni} the fifth-force here as ${F_5} = -
v_{\rm tg}^{} M_{\rm Pl}^{ - 1}{\nabla ^r}\phi/2  $. In this
regard, utilizing ${\nabla ^r}\phi  = {g^{rr}}\phi ' = {e^{ -
b}}\phi '$ and using relations (\ref{phi prime}), (\ref{grr})
and~(\ref{alphaBeta}), we obtain the fifth-force in the presented
chameleon gravity to be
\begin{equation}
{F_5} \approx\frac{{ v_{\rm tg}^{2} \, }}{2\,r}e^{-2v_{\rm
tg}^{2}}.
\end{equation}
Therefore, in this chameleon gravity, the coupling parameter
between the chameleon scalar field and the matter field gives rise
to the fifth-force. However, since the chameleon scalar field
model invokes a screening mechanism, it remains consistent with
the tests of gravity on small scales. This means that in
high-density regions, such as inside galaxies, the chameleon
scalar field is massive and suppressed, and the model behaves
effectively like the standard GR. Hence, the resulting fifth-force
has a very short-range and cannot be detected. However, in
low-density environments, such as galactic halos, the chameleon
scalar field becomes lighter, resulting in the emergence of a
long-range fifth-force that enhances gravity. Thus, this
additional gravitational force makes the rotation curves in the
outer regions of galaxies to be flat, as observed. Hence, instead
of relying on the concept of dark matter, the scalar field in the
chameleon gravity strengthens the gravitational interactions in
the halo regions and maintains the high orbital velocities of
stars. Therefore, the fifth-force in the chameleon gravity
provides a suitable alternative to dark matter and an explanation
for the dynamics of galactic halos.

At this stage, let us make the behavior of the presented chameleon
gravity clearer. In this regard, by equating relations
(\ref{speed}) and (\ref{speed2}) with each other and utilizing
relations (\ref{alphaBeta}) and (\ref{R0}), we obtain ${M^{\left(
\phi \right)}} \propto 4{\beta ^2}G^{ - 1}r $. Then, using
relation $ {M^{\left( \phi \right)}} = \int {4\pi {\rho ^{\left(
\phi \right)}}} {r^2}dr $, the chameleon scalar field density is
given as
\begin{equation}\label{new rho phi}
{\rho ^{\left( \phi  \right)}} \propto \frac{{{\beta ^2}}}{{\pi
G{r^2}}},
\end{equation}
which is more practical than relation \eqref{density2}. This
relation indicates that in the region of galactic halos, the mass
density of the chameleon scalar field decreases with the inverse
square of the radius. This is significant in understanding the
role of the chameleon scalar field in the dynamics of galaxies in
this region. In this regard, we have compared the obtained results
of the model with the corresponding results of the $ \Lambda CDM $
model with the Navarro-Frenk-White (NFW) halo profile introduced
in Ref.~\cite{Navarro}. Accordingly, we have plotted the
tangential speed diagrams for the NGC~$5533$ galaxy for both
models in Figure~$(2)$ using relation~\eqref{new rho phi} and
assuming that the mass density of matter in the visible region
remains constant. This figure indicates that the presented
chameleon gravity is closer to observations for smaller coupling
coefficients.
\begin{figure}[h]
\begin{center}
\includegraphics[scale=0.4]{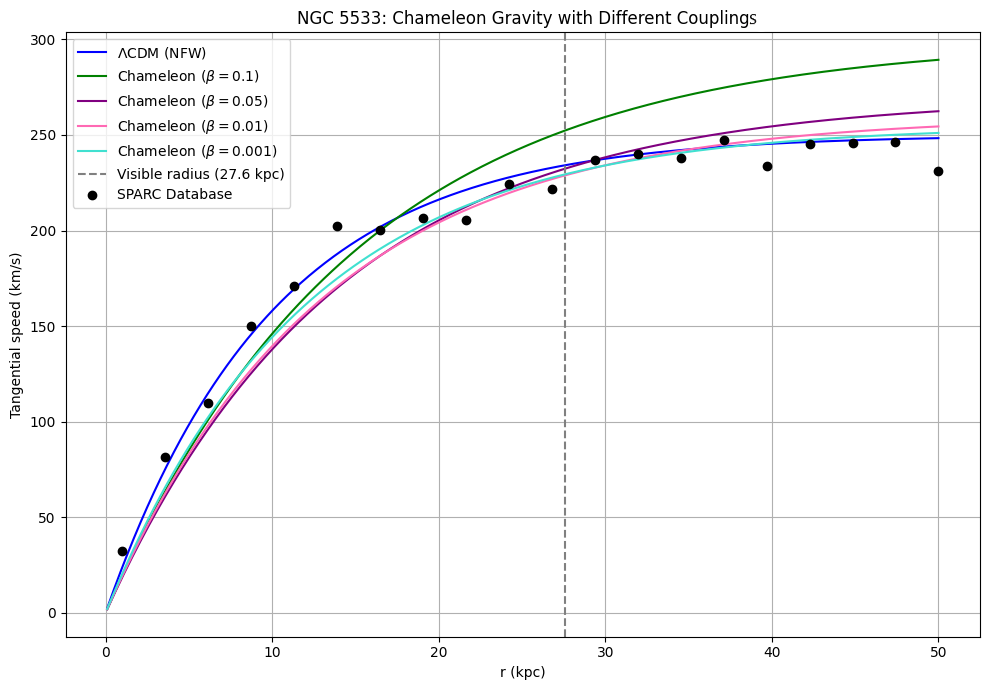}
\caption{The figure illustrates the tangential speed of the
NGC~$5533$ galaxy with data of the SPARC database.}
\end{center}
\end{figure}

Also, to compare the presented chameleon gravity with the $
\Lambda CDM $ model with the NFW halo profile\footnote{In the NFW
halo profile, we have assumed $ {r_s} = 0.15{R_{200}}.$},
 we have generated the following table for several galaxies
using the SPARC database. In this table, we have used
relation~(\ref{new rho phi}) to obtain the galactic halo mass for
the presented chameleon gravity. As shown, the values obtained
from the presented chameleon gravity closely align with the
corresponding values of the $ \Lambda CDM $ model with the NFW
halo profile.
\begin{center}
\boxed{\begin{array}{*{20}{c}}
  {\boxed{Galaxy}}&{\begin{array}{*{20}{c}}
  v_{\rm tg}^{} \\
  {\underline {km\,{s^{ - 1}}} }
\end{array}}&{\begin{array}{*{20}{c}}
  {{r_{\rm D}^{}}} \\
  {\underline {kpc} }
\end{array}}&{\begin{array}{*{20}{c}}
  {M_{\rm SPARC}^{\left( {m} \right)}} \\
  {\underline {{{10}^{11}}{M_ \odot }} }
\end{array}}&{\begin{array}{*{20}{c}}
  {M_{\rm NFW}^{\left( {\rm halo} \right)}} \\
  {\underline {{{10}^{11}}{M_ \odot }} }
\end{array}}&{\begin{array}{*{20}{c}}
  {M_{\rm chameleon}^{\left( \phi  \right)}} \\
  {\underline {{{10}^{11}}{M_ \odot }} }
\end{array}} \\
  {NGC\,5533}&{275}&{125}&{0.5}&{1.8}&{1.7} \\
  {NGC\,6946}&{180}&{110}&{0.4}&{7.8}&{7.8} \\
  {M\,33}&{135}&{90}&{0.1}&{3.5}&{3.7} \\
  {NGC\,925}&{110}&{90}&{0.1}&{2.4}&{2.4} \\
  {NGC\,2841}&{310}&{175}&{0.8}&{38.0}&{38.3} \\
  {NGC\,2903}&{190}&{115}&{0.3}&{9.2}&{9.3} \\
  {NGC\,7331}&{250}&{175}&{0.7}&{24.5}&{24.0}
\end{array}}
\end{center}

\section{Light-Deflection Angle}
\indent

In this section, we investigate the light-deflection angle as an
effect of DM for the chameleon gravity. To perform this
calculation, we utilize metric (\ref{metric}) in the region of
flat galactic rotation curves. The deflection angle
is~\cite{Weinberg}
\begin{equation}
\Delta\varphi = 2\left| {\varphi \left( {{r_0^{}}} \right) -
\varphi \left( \infty  \right)} \right| - \pi ,
\end{equation}
where $ r_{0}^{} $ is the radius of the closest approach to the
center of galaxy. The geodesic equation (\ref{geodesic}) for a
photon reduces to~\cite{Weinberg}
\begin{equation}\label{angle}
\varphi \left( {{r_0^{}}} \right) - \varphi\left( \infty  \right)
= {\int_{{r_0^{}}}^{\infty} e^{b(r)/2} {\left[ {{{ \left(
{\frac{r}{{{r_0^{}}}}} \right)}^2}{e^{a\left( {{r_0^{}}} \right) -
a\left( r \right)}} - 1} \right]} ^{ - \frac{1}{2}}}\frac{{dr}}{r}
\end{equation}
with $ r_{0}^{} $ in the region of flat galactic rotation curves
as $ s \leq r_0^{} \leq {r_{\rm D}^{}} $. Hence, due to the
obtained solutions, we have to split the above integral into two
parts, i.e. from $r_0^{}$ to $r_{\rm D}^{}$ and from $r_{\rm
D}^{}$ to $\infty$. Then, utilizing relations (\ref{gtt}) and
(\ref{grr}) for the first integral and $e^{a\left( r
\right)}=e^{-b\left(r \right)}$ with relation (\ref{e^-b}) for the
second integral behind the region of galactic halos, we obtain
\begin{equation}\label{delta phi}
\varphi \left( {{r_0^{}}} \right) - \varphi \left( \infty
\right)\! =\! e^{v_{\rm tg}^{2}} \!\int_{{r_0^{}}}^{{r_{\rm
D}^{}}}
{{{\!\left[ {{{\!\left( {\frac{{{r_0^{}}}}{r}} \right)}^{v_{\rm tg}^{2}  - 2}}\! -\! 1}
\right]}^{ - \frac{1}{2}}}} \frac{{dr}}{r} \hfill \\
+\! \int_{{r_{\rm D}^{}}}^\infty  {\!\frac{1}{{\sqrt {1\! -\!
\frac{{M^{\left( \rm tot  \right)}}}{{4\pi r M_{\rm Pl}^2}}} }}}
   {\left[ {\!\frac{{1\! -\! \frac{{M^{\left( \rm tot  \right)}}}{{4\pi {r_0^{}} M_{\rm Pl}^2}}}}{{1
    \!-\! \frac{{M^{\left( \rm tot  \right)}}}{{4\pi r M_{\rm Pl}^2}}}}{{\left( {\frac{r}{{{r_0^{}}}}} \right)}^2}
    \!-\! 1} \right]^{ - \frac{1}{2}}}\!\frac{{dr}}{r}.
\end{equation}
Therefore, from relation (\ref{delta phi}), we get
\begin{eqnarray}\label{Delta Phii}
\Delta \varphi  =2\!\!\!\!\!\!\!\!\!\! && \Bigg|\, {e^{v_{\rm
tg}^{2}} \left( {\frac{2}{{2 - v_{\rm tg}^{2} }}} \right)\arctan
\left[ {\sqrt {{{\left( {\frac{{{r_0^{}}}}{{{r_{\rm D}^{}}}}}
\right)}^{v_{\rm tg}^{2} - 2}} - 1} } \right]} \cr
  && + \arcsin \left(
{\frac{{{r_0^{}}}}{{{r_{\rm D}^{}}}}} \right) + \frac{{M^{\left(
\rm tot \right)}}} {{8\pi {r_0^{}}M_{\rm Pl}^2}}\left[ {2 - \sqrt
{1 - {{\left( {\frac{{{r_0^{}}}}{{{r_{\rm D}^{}}}}} \right)}^2}} -
\sqrt {\frac{{{r_{\rm D}^{}} - {r_0^{}}}}{{{r_{\rm D}^{}} +
{r_0^{}}}}} } \right] \Bigg|
 - \pi.
\end{eqnarray}
The first term in the above expression represents the effect of
the presented chameleon gravity on the light-deflection angle due
to the region of galactic halos. When $ r_{0}^{} = r_{\rm D}^{} $,
this term vanishes and the deflection angle originates only from
the baryonic matter. In this regard, the remaining terms, i.e. the
second and third terms, of Eq. (\ref{Delta Phii}) yield $\Delta
\varphi\mid_{r_{0}^{} = r_{\rm D}^{}} ={M^{\left( \rm tot
\right)}}/(2\pi\,r_{\rm D}^{} M_{\rm Pl}^2) = 4G{M^{\left( \rm tot
\right)}}/r_{\rm D}^{}$, which is consistent with the GR
predictions in, e.g., Refs.~\cite{Sefiedgar,ZareFarhoudi,Y.
Sobouti}. In Figure~$(3)$, we have plotted the first term as a
function of $r_0^{}/r_{\rm D}^{}$ in the region of flat rotation
curves for the NGC~$5533$ galaxy with the tangential speed of the
test particle $275\ km/s$~\cite{Y. Sobouti}, hence $ {v_{\rm
tg}^{}}\approx 9.17 \times 10^{-4}$ in the natural units. Note
that in drawing this figure, we have ignored the insignificant
value of the coefficient of the first term due to the very small
value of $v_{\rm tg}^{} $.
\begin{figure}[h]
\begin{center}
\includegraphics[scale=0.6]{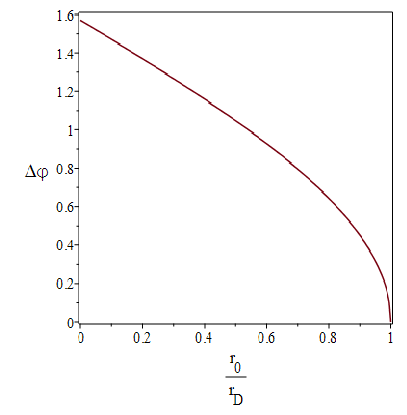}
\caption{The figure illustrates the effect of the added term
resulting from the presented chameleon gravity on the
light-deflection angle as a function of $ r_{0}^{}/r_{\rm D}^{}$
for the NGC~$5533$ galaxy with $ {v_{\rm tg}^{}}\approx 9.17
\times 10^{-4}$ in the natural units.}
\end{center}
\end{figure}
Comparing the deflection angle obtained from the presented
chameleon gravity with other modified gravitational theories, such
as the $ f(R) $ gravity~\cite{Sefiedgar}, the $ f(R,T) $ gravity
and a generalization of the pseudoisothermal DM
model~\cite{ZareFarhoudi,Gunn}, and the symmetron field
model~\cite{ZareHonar}, indicates that they all exhibit similar
behavior, in that the deflection angle in the region of galactic
halos decreases as the radius of galaxy increases.

\section{Radar Echo Delay}
\indent

The time takes for light to travel around a massive celestial body
is longer than the corresponding time calculated by Newtonian
gravity. This phenomenon is the result of gravitational pull of
massive objects, such as black holes, and can be identified
through the use of radar echo delay. Physicist Irwin Shapiro was
the first to observe this delay and proposed a method for
measuring it~\cite{Shapiro}. The time required for light to go
from $ r_{c} $  to $ r $ or from $ r $ to $ r_{c} $,
is~\cite{Weinberg}
\begin{equation}\label{delay}
t\left( {r,{r_c}} \right) = {\int_{{r_c}}^r {\frac{{{e^{b\left( r
\right)/2}}}}{{{e^{a\left( r \right)/2}}}}\left[ {1 -
\frac{{{e^{a\left( r \right)}}}}{{{e^{a\left( {{r_c}}
\right)}}}}{{\left( {\frac{{{r_c}}}{r}} \right)}^2}} \right]} ^{ -
\frac{1}{2}}}dr,
\end{equation}
where $ r_{c} $ represents the closest approach to the celestial
body. For the presented chameleon gravity, to determine the time
delay, we consider that $ r $ is located outside the region formed
by the galaxy, while $ r_{c} $ is within the region of flat
rotation curves. Hence, due to the obtained solutions, we have to
split integral (\ref{delay}) into two parts, i.e. from $r_c^{}$ to
$r_{\rm D}^{}$ and from $r_{\rm D}^{}$ to $r$. Then, utilizing
relations (\ref{gtt}) and (\ref{grr}) for the first integral and
$e^{a\left( r \right)}=e^{-b\left(r \right)}$ with relation
(\ref{e^-b}) for the second integral behind the region of galactic
halos, we obtain
\begin{eqnarray}
t\left( {r,{r_c}} \right) =\!\!\!\!\!\!\!\!\!\! && e^{2v_{\rm
tg}^{2}}\int_{{r_c}}^{{r_{\rm D}^{}}} {\sqrt {{{\left(
{\frac{{s}}{{ r}}} \right)}^{v_{\rm tg}^{2}} }} } {\left[ {1 -
{{\left( {\frac{r}{{{r_c}}}} \right)}^{v_{\rm tg}^{2}  - 2}}}
\right]^{ - \frac{1}{2}}}dr\cr
 && + \int_{{r_{\rm D}^{}}}^r {\frac{1}{{1 -
\frac{{{M^{\left( {\rm tot} \right)}}}}{{4\pi\, r M_{\rm
Pl}^2}}}}} \left[ {1 - \frac{{1 - \frac{{{M^{\left( {\rm tot}
\right)}}}}{{4\pi\, r M_{\rm Pl}^2}}}}{{1 - \frac{{{M^{\left( {\rm
tot} \right)}}}}{{4\pi\,{r_c} M_{\rm Pl}^2}}}}{\left(
{\frac{{{r_c}}}{r}} \right)}^2} \right]^{ - \frac{1}{2}}dr.
\end{eqnarray}
By solving the first term exactly through the change variable $
{\left( r/r_c \right)^{v_{\rm tg}^{2}  - 2}} \equiv {\sin
^2}\theta $ and using the Robertson expansion (see
Refs.~\cite{Sefiedgar,Weinberg}) to solve the second term, we
obtain
\begin{eqnarray}
t\left( {r,{r_c}} \right) \approx \!\!\!\!\!\!\!\!\!\! &&
{e^{2v_{\rm tg}^{2}}\sqrt {s^{v_{\rm tg}^{2} }} } \left(
\frac{{2r_c^{1 - \frac{v_{\rm tg}^{2} }{2}}}}{{2 - v_{\rm tg}^{2}
}}\right) \cot \left[ {\arcsin {{\left( {\frac{{{r_c}}}{{{r_{\rm
D}^{}}}}} \right)}^{1-\frac{v_{\rm tg}^{2} }{2}}}} \right] \cr
 && + \sqrt {{r^2} - r_{\rm D}^{2}}+ \frac{{{M^{\left(
{\rm tot} \right)}}}}{{4\pi M_{\rm Pl}^2}} \ln \left( {\frac{{r +
\sqrt {{r^2} - r_{\rm D}^{2}} }}{{{r_{\rm D}^{}}}}} \right) +
\frac{{{M^{\left( {\rm tot} \right)}}}}{{8\pi M_{\rm Pl}^2}}\sqrt
{\frac{{r - {r_{\rm D}^{}}}}{{r + {r_{\rm D}^{}}}}}.
\end{eqnarray}
The first term in the above expression represents the impact of
the presented chameleon gravity due to the region of flat rotation
curves, while the other terms exactly account the time delay for
the influence of GR in the absence of the region containing flat
rotation curves, see, e.g., Ref.~\cite{Weinberg}. The leading term
$\sqrt {{r^2} - r_{\rm D}^{2}} $ represents the expected behavior
of light traveling in a straight line at a constant speed. Also,
the first term decreases with increasing radius $r_c$ towards the
end of the galactic halo. It is obvious that in the absence of the
flat rotation curves, i.e. when $r_c=r_{\rm D}^{}$, the first term
vanishes and only the time delay for GR remains.

\section{Conclusions}\label{Sec7}
\indent

In standard chameleon gravities, a scalar field has been
introduced to evade constraints on the equivalence principle
violation in laboratory experiments. Such a scalar field
non-minimally interacts with the baryonic matter field, and
usually plays the role of dark energy in scalar-tensor theories to
explain the accelerated expansion of the universe. In the
chameleon gravity presented in this work, we have investigated the
ability of this scalar field to act also as DM in addition to dark
energy. Then, by obtaining the expression for the tangential speed
in the region of galactic halos, we have applied this approach to
explain the issue of flat galactic rotation curves. Moreover, we
have obtained the effects on the angle representing the deflection
of light and the time representing the radar echo delay.

To perform this task, we have considered the chameleon gravity for
a static spherically symmetric spacetime that has the components
of the metric with respect to the energy-momentum tensor of the
matter field and the chameleon scalar field. We have also assumed
that the matter field is non-relativistic baryonic matter with
zero equation of state, and that the chameleon scalar field
inherits the spacetime symmetry and depends only on the radial
coordinate. Hence, the conservation equation of the
energy-momentum tensor of the matter field dictates that the
matter density in both the Jordan and Einstein frames cannot
depend on the time. In this vein, we have assumed that those also
depend only on the radial coordinate.

Next, we have focused on the region of galactic halos, where the
baryonic matter density is negligible compared to the mass density
associated with the chameleon scalar field, while the baryonic
matter dominates in the interior of galaxies. Accordingly, in the
region of galactic halos, we have obtained that the mass
associated with the chameleon scalar field varies linearly with
the radius of galaxy. Furthermore, the obtained result indicates
that the tangential speed is constant in that region consistent
with observational data. In fact, we have found that the mass
associated with the chameleon scalar field explains the
observation of flat galactic rotation curves without any need to
introduce a mysterious DM, i.e., it plays the role of the mass
associated with DM. In other words, since finding the nature and
properties of DM is one of the outstanding questions in
contemporary cosmology, such a result represents a different
approach and interestingly demonstrates the capability of the
presented chameleon gravity and allows for investigating its
validity.

In the continuation, we have considered solutions in the region of
galactic halos that differ only slightly from classical GR, but
have similar symmetry behind this region. Hence, we have first
obtained that the chameleon scalar field varies proportionally to
the logarithm of the radius of galaxy, which we had assumed
vanishes at the boundary radius of the region dominated by the
baryonic matter. We have then achieved the metric components and
have shown that the presented chameleon gravity has a well-defined
Newtonian limit and can describe the geometry of spacetime in the
region of flat galactic rotation curves. Meanwhile, in this
region, the presented chameleon gravity dictates that the
chameleon coupling parameter can only have positive values,
although much less than $1$, which is well consistent with the
expected magnitude and possible observational bounds, and that the
power $n$ of the chameleon scalar field in the standard chameleon
potential must have odd values. We have also shown that in the
region of galactic halos, the mass density of the chameleon scalar
field decreases with the inverse square of the radius. Then, using
the SPARC database, we have compared the obtained results with the
corresponding ones of the $ \Lambda CDM $ model with the NFW halo
profile for several galaxies. The outcomes show that the results
of the model closely align with the corresponding results of the $
\Lambda CDM $ model with the NFW halo profile.

Moreover, by considering a test particle moving in a timelike
geodesic of the static spherically symmetric metric that differs
only slightly from classical GR, we have obtained the fifth-force,
which varies proportionally to the inverse of the radius of
galaxy. The fifth-force is one of the most significant features of
scalar-tensor theories of gravity. In the chameleon cosmology, it
arises notably when the chameleon scalar field is coupled
conformally to the matter sector while remaining minimally coupled
to the metric. However, due to the screening mechanism of
chameleon gravities, such a fifth-force can be suppressed to pass
existing tests of gravity at small scales.

Finally, we have studied the light-deflection angle and the radar
echo delay as two effects of DM in the presented chameleon
gravity. In this regard, we have shown that as the radius of
galaxy increases, the effect on the angle of light passing through
the region around galactic halos decreases. This effect is similar
to the corresponding results obtained in other modified
gravitational theories. Also, in the absence of the above effect,
the remaining terms are just the GR predictions. Furthermore, we
have obtained the impact of the presented chameleon gravity on the
time delay of photons passing through the region of galactic halos
as an additional term to the exact terms of the influence of GR.
The result obtained indicates that this time delay decreases with
increasing radius towards the end of the galactic halo.


\end{document}